\documentclass[smallextended,numbook]{svjour3}

\journalname{Journal of Statistical Physics}
\journalname{}

\usepackage{ulem}

\usepackage{graphicx,float,amsmath,amssymb,mathrsfs}

\usepackage[english,francais]{babel}

\usepackage{url,hyperref}  

\usepackage[usenames,dvipsnames]{color}

\begin{document}

\catcode`@=11 
\renewcommand\tableofcontents{%
  \section*{\contentsname}%
  \@starttoc{toc}%
}
\catcode`@=12

\selectlanguage{english}

\title{Stress Response of Granular Systems}

\author{Kabir Ramola \and Bulbul Chakraborty}

\institute{
K. Ramola
\at 
Martin Fisher School of Physics, Brandeis University, Waltham, MA 02453, USA\\
\email{kramola@brandeis.edu}
\and
B. Chakraborty
\at
Martin Fisher School of Physics, Brandeis University, Waltham, MA 02453, USA\\
\email{bulbul@brandeis.edu}
}

\date{\today}

\maketitle

\vspace{-0.5cm}
\begin{abstract}
We develop a framework for stress response in two dimensional  granular media, with and without friction,  that respects vector force balance at the microscopic level. We introduce local gauge degrees of freedom that determine the response of contact forces between constituent grains on a given, disordered, contact network, to external perturbations.
By mapping this response to the spectral properties of the graph Laplacian corresponding to the underlying contact network, we show that this naturally leads to {\it spatial localization} of forces. We present numerical evidence for localization using exact diagonalization studies of network Laplacians of soft disk packings. Finally, we discuss the role of other constraints, such as torque balance, in determining the stability of a granular packing to external perturbations. 
  
\end{abstract}



\tableofcontents 

\section{Introduction}

Force transmission in granular materials is a well-studied problem with wide ranging applications \cite{behringer_review_2016,majumdar_coppersmith_science_1995,majumdar_coppersmith_pre_1996,mueth_jaeger_pre_1998,cates_wittmer_prl_1998,bouchaud_claudin_crp_2002,geng_prl_2001,erpelding_epl_2010}
and several properties of stress transmission in granular media continue to be the subject of active research.
Several recent studies have addressed the question of the stress response of granular packings \cite{goldenberg_prl_2002,goldenberg_nature_2005,goldenberg_prl_2006,goldenberg_pre_2008,yan_arxiv,reydellet_Clement_PRL2001,atman_epje_2005,atman_jphyscm_2005,atman_epl_2013,gland_epje_2006},
showing that the usual elasticity theories for homogeneous materials do not apply to materials with granular constituents, and therefore new frameworks need to be developed to deal with such systems.  
One of the most important characteristics that has emerged from these studies is the inhomogeneous nature of this stress propagation \cite{bouchaud_epje_2001,socolar_epje_2002}. A striking aspect of granular systems is that forces are primarily carried by a sparse, tenuous network of contacts that have become known  as ``force chains". Experiments using photoelastic beads provide clear evidence of this phenomenon \cite{geng_prl_2001}, and force chains have emerged as the defining characteristic of granular solids. Yet, at present we do not have a theory of how forces are localized in space and the role played by network disorder on the character of this spatial localization \cite{geng_prl_2001}.

Static granular media is controlled by the constraints of mechanical equilibrium, which is the main ingredient in models of such systems.
Several theoretical frameworks have been proposed to explain how granular materials respond to external forces \cite{bouchaud_review_2002}. These include lattice-based models such as the $q$-model 
\cite{majumdar_coppersmith_science_1995,majumdar_coppersmith_pre_1996}, and its extensions \cite{narayan_pre_2000,tighe_pg_2009}, and continuum models that posit some history-dependent relation between the components of the stress tensor \cite{bouchaud_cates_claudin_jpI_1995,vanel_howell_pre_1999}. These frameworks lead to Partial Differential Equations (PDEs) that are elliptic, hyperbolic or parabolic depending on these closure relations \cite{bouchaud_claudin_crp_2002}. 
The $q$-model, which incorporates scalar force balance in a model of granular piles in a gravitational field \cite{majumdar_coppersmith_science_1995,majumdar_coppersmith_pre_1996}, successfully accounts for the distribution of contact forces in the large force limit. In the continuum limit, the model reduces to the diffusion equation, predicting a horizontal spread of force-bearing contacts that grows as the square-root of the depth in the granular pile.  Such a spread is observed in experiments under certain conditions of preparation \cite{geng_prl_2001}. Models with hyperbolic PDEs and disorder predict a wave-like propagation, which is similar to experimental observations in ordered arrays of grains \cite{geng_prl_2001}. Thus, the relationship between the stress response of granular materials and the underlying disorder of the contact network remains to be understood.

One of the important open questions in the field of granular systems is, how does one account for {\it vector force balance} on a disordered network of contacts?
In this paper we address this problem by studying the response of a two dimensional granular material to an externally applied force.
We show that the inhomogeneous propagation of stress at the grain level can be linked to the inherent randomness in the underlying fabric of contacts.
We connect the problem of stress transmission in this system to that of diffusion on the disordered planar graph formed by the contacts between the constituent grains. This allows us to develop a theoretical framework to describe the response of such a granular system to an imposed body force, and thereby probe the origin of the force localization properties in such systems. In this work, we do not address the question of what the proper continuum level description is of stress propagation in granular media, rather, we demonstrate that the response at the granular scale is controlled by the spectral properties of the graph Laplacian describing the disordered contact network.

The constraints of mechanical equilibrium, that are necessarily satisfied in a static granular packing, lead to a gauge potential formulation of forces and stresses \cite{satake_mom_1993,ball_blumenfeld_prl_2002}. We show that the response of the internal forces to an applied body force can be described by an additional set of gauge potentials, which satisfy equations involving the Laplacian of the contact network. By introducing these auxiliary fields we can account for the change in the local stress tensor induced by an external perturbation. In addition to introducing the formalism, we present results of exact diagonalization studies of network Laplacians produced using numerical simulations of soft disk packings. These clearly illustrate the localization properties of force transmission in such materials. Although our formulation is valid for a general frictional granular packing, in our simulations we focus specifically on linear spring potentials of fricitionless soft disks. Finally we discuss the role of additional constraints such as torque balance in determining the stability of granular packings to external perturbations.

\section{Local Constraints and Gauge Potentials}

In this section we discuss the constraints that need to be satisfied at the local level in a static granular packing. The local nature of these constraints allows a description of the system in terms of gauge potentials. Local constraints are also crucial in determining the response of jammed packings to external perturbations, and give rise to deviations from linear elasticity. 

\begin{figure}
\begin{center}
\includegraphics[width=0.55\linewidth]
{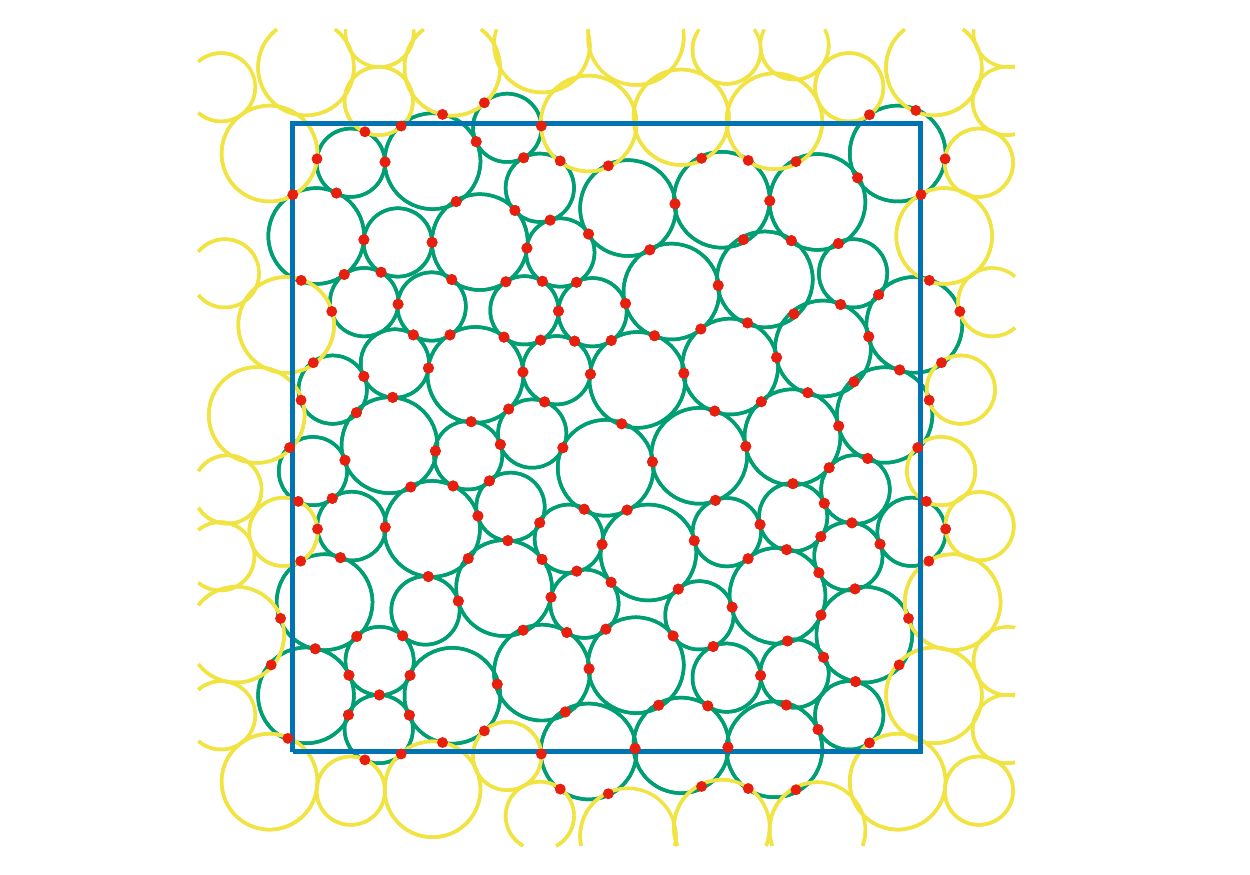}
\hspace{-2cm}
\includegraphics[width=0.55\linewidth]
{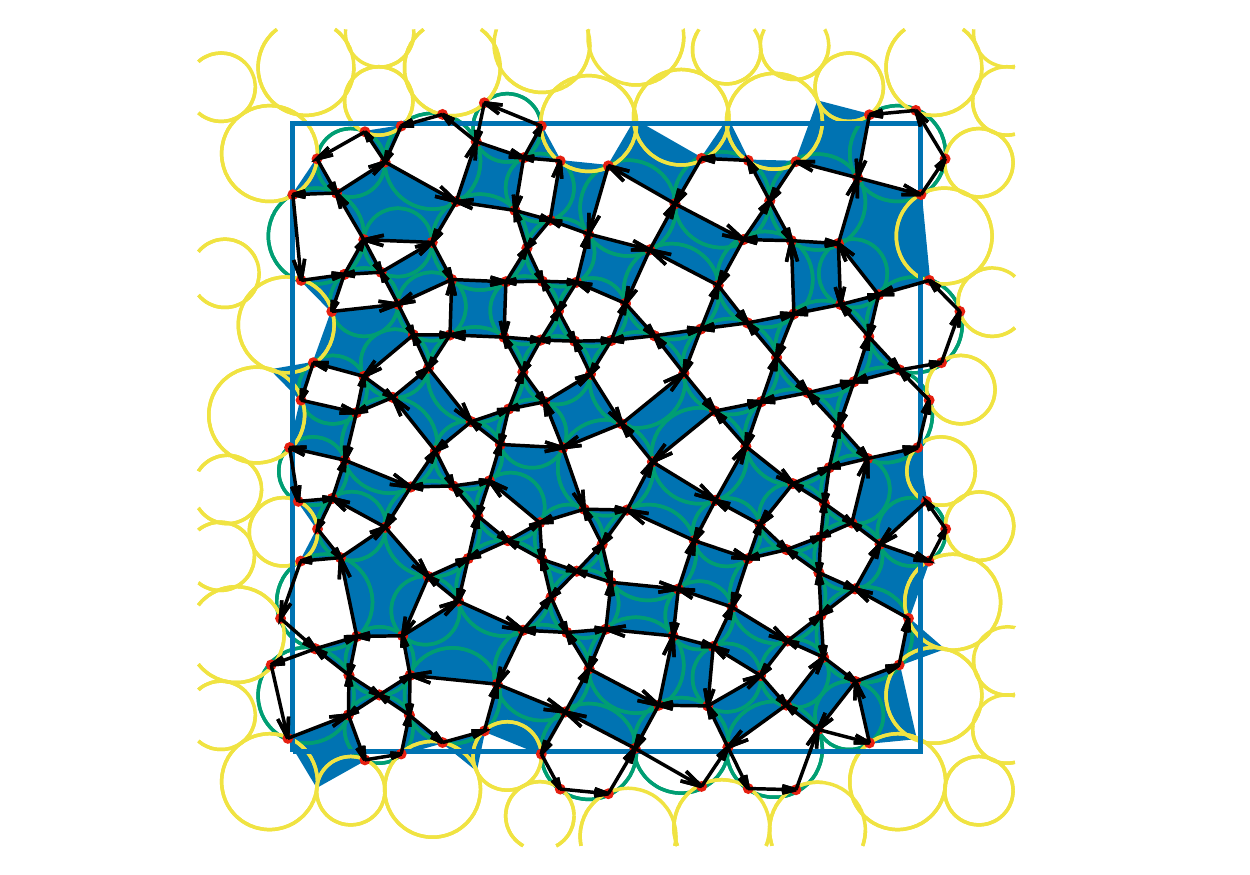}
\caption{({\bf Left}) A jammed packing of bidispersed frictionless disks with periodic boundary conditions. The contacts between the grains are idealized as points (red dots), and display a spatially disordered structure. ({\bf Right}) The same configuration with the associated grain polygons (white) and void polygons (blue). The grain polygons are formed by connecting the contact points within each grain in a cyclic manner. The void polygons are formed by cyclically connecting the contacts associated with each void. The grain and the void polygons together tessellate the entire space. The adjacency graphs formed by the two networks (grains and voids) are dual to each other.}
\label{grain_void_figure}
\end{center}
\end{figure}

\subsection{Grains and Voids in Jammed Configurations}
Granular systems are inherently porous with voids interspersed between grains in {\it contact}, and are inhomogeneous on the granular scale. 
This non-isotropic fabric of the underlying material leads to interesting properties of stress transmission between the grains in the system.
A challenge in granular mechanics is to understand how this ``graininess" \cite{kadanoff_review_rmp_1999} affects the bulk behavior of granular solids. 
In order to associate local quantities to granular packings with complex internal structures such as non-convex voids (see Fig. \ref{grain_void_figure}), it becomes necessary to associate well defined regions of space to specific parts of the packing. In two dimensional granular packings, the plane can be decomposed into polygonal regions belonging to {\it grains} as well as to {\it voids} (see Fig. \ref{grain_void_figure}).  The grain polygons are formed by connecting the contact points on the boundaries of each grain in a cyclic manner. The void polygons are formed in a similar manner by cyclically connecting the contacts associated with each void. The grain and void polygons together tessellate the entire space. This construction allows a decomposition of the space into well defined polygonal regions and provides a way of probing the spatial structure of granular materials \cite{ramola_chakraborty_jstat_2016,ramola_chakraborty_arxiv_2016}.
The two graphs formed by adjacent grains and adjacent voids are {\it dual} to each other. As we show in the next sections, this construction allows one to construct local gauge fields associated with these polygons which encodes the local force balance conditions.

\subsection{Force Balance}
The stability of granular materials stems from the fact that each individual  configuration is in {\it mechanical equilibrium}. The internal stresses in jammed packings are mediated via {\it contact forces} between the constituent grains which are pairwise within the system. The mechanical equilibrium condition translates to the fact that the forces acting on every grain sum to zero. For packings with only contact forces within the system, this condition can be represented by the equation 
\begin{equation}
\sum_{c} \vec{f}_{g,c} = 0,
\label{force_sum_zero}
\end{equation}
where $\vec{f}_{g,c}$ represents the force acting on the grain $g$, through the contact $c$, including both normal and tangential, frictional forces. The sum is taken over all the contacts $\{c\}$ for a given grain $g$. Next, Newton's {\it third law} dictates that 
\begin{equation}
\vec{f}_{g,c} = -\vec{f}_{g',c},
\label{third_law}
\end{equation}
at each contact $c$ between the grains $g$ and $g'$.
These two equations (Eqs. (\ref{force_sum_zero}) and (\ref{third_law})) can be used to construct an alternative representation of the forces in a two dimensional granular packing in terms of vector height fields which we discuss in the next section. 

\begin{figure} 
\begin{center}
\includegraphics[width=0.6\linewidth]
{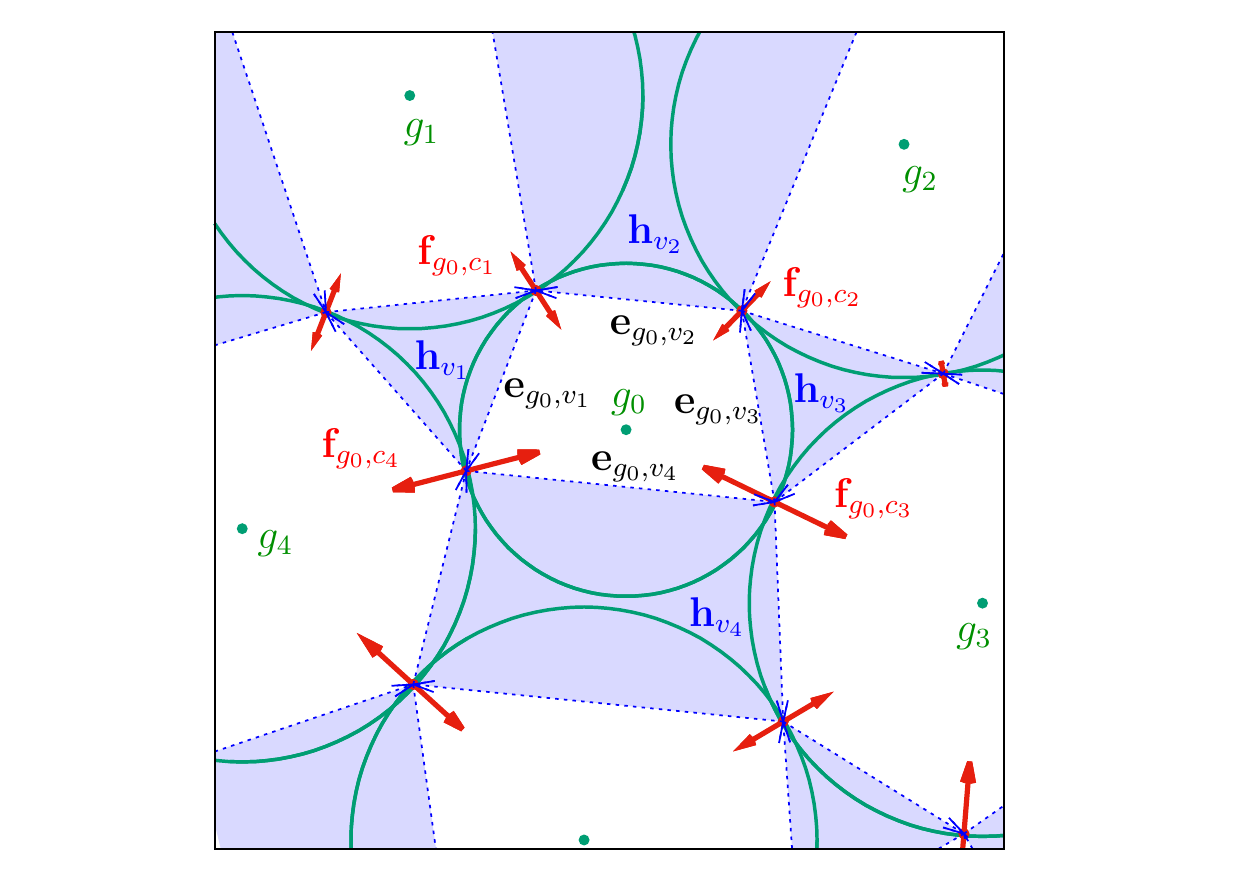}
\caption{The height fields $\{\vec{h}\}$ are associated with the void polygons $\{v\}$ (shaded light blue). The grain polygons $\{g\}$ are shown in white. The forces between the grains are represented by (bidirectional) arrows with $\vec{f}_{g,c} = -\vec{f}_{g',c}$ at each contact $c$. The forces at each contact are given by the difference of heights on the two voids associated with each contact (with a cyclic convention, i.e. $\vec{f}_{g_0,c_1} = \vec{h}_{v_1} - \vec{h}_{v_2}$). The vectors $\vec{e}_{g,v}$ define the vectors connecting the two contacts on grain $g$ that form an edge of the void polygon $v$.}
\label{explanatory_figure1}
\end{center}
\end{figure} 


\subsection{Height Fields}
\label{height_section}
The force balance condition combined with Newton's third law naturally leads to an alternative representation of the forces in the system, parametrized by a vector height field that lives on the edges, or the faces of the void polygons \cite{satake_mom_1993,ball_blumenfeld_prl_2002,henkes_chakraborty_pre_2009,degiuli_thesis,degiuli_pre_2011}. In this work we choose to place these fields on the faces of the void polygons (see Fig. \ref{explanatory_figure1}). As seen from Fig. \ref{explanatory_figure1}, each contact in a two dimensional granular packing is shared by two void polygons. We define a set of height vectors on the voids $\lbrace \vec {h}_{v} \rbrace$, where $v$ indexes the voids in the system. The force at each contact $c$ is given by the {\it difference of heights} on the two voids associated with each contact
\begin{equation}
\vec{f}_{g,c} = \vec{h}_{v'} - \vec{h}_{v},
\label{height_definition}
\end{equation}
with $v$ and $v'$ ordered cyclically around the centre of the grain $g$. This naturally leads to Eq. (\ref{third_law}), as the force on grain $g'$ through the contact $c$ is simply $\vec{f}_{g',c} = \vec{h}_{v} - \vec{h}_{v'}$.
It is interesting to note that although the contact forces are associated with the network of contacts between grains, the height vectors are {\it associated with the dual lattice} of the contact network i.e. the network of voids $\{v\}$.
The cyclic convention allows one to visualize the heights as vector currents circulating in the counterclockwise direction on the edges of a void polygon \cite{ball_blumenfeld_prl_2002}.
Given a set of contact forces, the definition of $\lbrace \vec {h}_{v} \rbrace$ is unique, modulo a choice of origin: they are gauge potentials for the stress tensor.
This uniqueness of the height representation is ensured by the force balance condition (Eq. \ref{force_sum_zero}). Traversing a path around any non-trivial loop within the system returns to the same value of the height field. This can be easily seen by circulating around each grain, which form the basic loops in the system. As an example we consider a grain $g_0$ which has four neighbours $g_{1,2,3,4}$ in contact through contacts $c_{1,2,3,4}$ (see Fig. \ref{explanatory_figure1}). Using the definition of the heights in Eq. (\ref{height_definition}), the forces acting on the grain $g_0$ are given by 
\begin{align}
\begin{tabular}{c c c}
    $\vec{f}_{g_0,c_1}$ & $=$ & $\vec{h}_{v_1} - \vec{h}_{v_2}$,\\
    $\vec{f}_{g_0,c_2}$ & $=$ & $\vec{h}_{v_2} - \vec{h}_{v_3}$,\\
    $\vec{f}_{g_0,c_3}$ & $=$ & $\vec{h}_{v_3} - \vec{h}_{v_4}$,\\
    $\underbrace{\vec{f}_{g_0,c_4}}_{\sum_c \vec{f}_{g_0,c}  = 0}$ & $=$ & $\underbrace{\vec{h}_{v_4} - \vec{h}_{v_1}.}_{\sum_{(v,v')} \vec{h}_{v} -  \vec{h}_{v'}  = 0}$\\  
  \end{tabular}
\label{uniqueness_of_heights}
\end{align}
The left hand side represents the force on every contact of the grain $g_0$ which sums to zero, while the right hand side represents the difference in the heights at each contact. The sum is taken over all adjacent pairs  of voids $(v,v')$ surrounding the grain $g_0$. Starting with a value $\vec{h}_{v_1}$ on the first void, the values of the heights around the grain $g_0$ can be cyclically constructed using the contact forces. The force balance condition thus necessitates that the heights around the loop return to the same value $\vec{h}_{v_1}$. The non-trivial nature of the loops that can occur in higher dimensions is the main obstacle in extending this simple height construction beyond two dimensions.

Given the forces within the system, the stress tensor for a given packing is then defined as
\begin{eqnarray}
\hat{\sigma} &= & \frac{1}{V}\sum_g \hat{\sigma}_g, \nonumber \\
\hat{\sigma}_g & = & \sum_{c} \vec{r}_{g,c} \otimes \vec{f}_{g,c} = \sum_{v} \vec{e}_{g,v} \otimes \vec{h}_{v}~.
\end{eqnarray}
Here $\vec{r}_{g,c} = \vec{r}_{c} - \vec{r}_{g}$, with $\vec{r}_{c}$ being the position of the contact $c$, and $\vec{r}_{g}$ being the position of the centre of the grain $g$. $V$ represents the volume of the entire system. The vectors $\vec{e}_{g,v}$ define the vectors connecting the two contacts on grain $g$ (cyclically) that form an edge of the void  polygon $v$ (illustrated in Fig. \ref{grain_void_figure}). Using the representation of $\hat{\sigma}_g$  in terms of the height vectors, it is easy to show that the total force moment tensor reduces to a boundary term \cite{henkes_chakraborty_pre_2009}, which is the discrete version of Stokes' theorem.
From a continuum perspective, the height fields can therefore be viewed as the gauge potential that enforces $\nabla \cdot \hat {\sigma} =0$ \cite{ball_blumenfeld_prl_2002}.

Finally it is important to note that a well defined height field at the local level can only be constructed in situations where pairwise forces are the only forces present in the system, since both Eqs. (\ref{force_sum_zero}) and (\ref{third_law}) are essential to the construction. Therefore several cases of interest are excluded from this framework, most notably systems where body forces, such as gravity, act on the grains. Similarly granular systems that are perturbed by an external force are also not amenable to a height description. It is therefore important to extend the gauge field treatment to such cases. In Section \ref{response_to_perturbation_section} we extend this height construction to such cases, which allows us to study the transmission of stress within two dimensional granular materials.

\subsection{Additional Constraints}
\label{additional_constraints_section}
In addition to local force balance, a mechanically stable configuration must also satisfy {\it torque balance} at the grain level. These are represented by the set of equations
\begin{equation}
\sum_{c} \vec{r}_{g,c} \times  \vec{f}_{g,c} = 0,
\end{equation}
where the sum is taken over all the contacts $\{c\}$ for a given grain $g$.  In continuum, the torque balance constraint translates to the symmetric property  of the stress tensor.
Finally, in order for the packing to be valid, it must also satisfy additional constraints in real space. For example, the vector sum of inter particle distances taken over every closed trajectory in the system, must sum to zero.  For the planar contact network in two dimensional packings, these constraints can be parameterized by the set of basic loops that enclose each void. These can then be represented by the set of equations
\begin{equation}
\sum_{(g,g')} \vec{r}_{g,g'} = 0,
\label{real_space_constraints}
\end{equation}
where $\vec{r}_{g,g'} = \vec{r}_{g'} - \vec{r}_{g}$ is the interparticle distance vector between two adjacent grains located at $\vec{r}_{g'}$ and $\vec{r}_{g}$ respectively. The sum is taken over all adjacent pairs $(g,g')$ surrounding each void $v$.
This illustrates the fact that for systems where the forces are related to the inter particle distances through a force law, the distrbution of stress in the system is intimately linked to the underlying real space network and its constraints.

\section{Response to a Perturbation} 
\label{response_to_perturbation_section}
In this section, 
we use the gauge potential formalism to analyse the response of granular packings to external perturbations such as an imposed body force {\it while respecting the local force balance constraint}.  In the presence of body forces, the continuum equation of mechanical equilibrium is   $\nabla \cdot \hat{\sigma} = -\vec{f}^{\rm body}$. We generalize the height field construction, which imposes mechanical equilibrium at the discrete grain level, to this situation.
At the granular level, the mechanical equilibrium condition with an imposed body force $\vec{f}^{\rm body}_{g}$ on each grain $g$ now becomes
\begin{equation}
\sum_{c} \vec{f}_{g,c} = -\vec{f}^{\rm body}_{g}.
\label{force_sum_body}
\end{equation}
with the sum taken over all the contacts $\{c\}$ of the grain $g$.

\begin{figure} 
\begin{center}
\includegraphics[width=0.6\linewidth]
{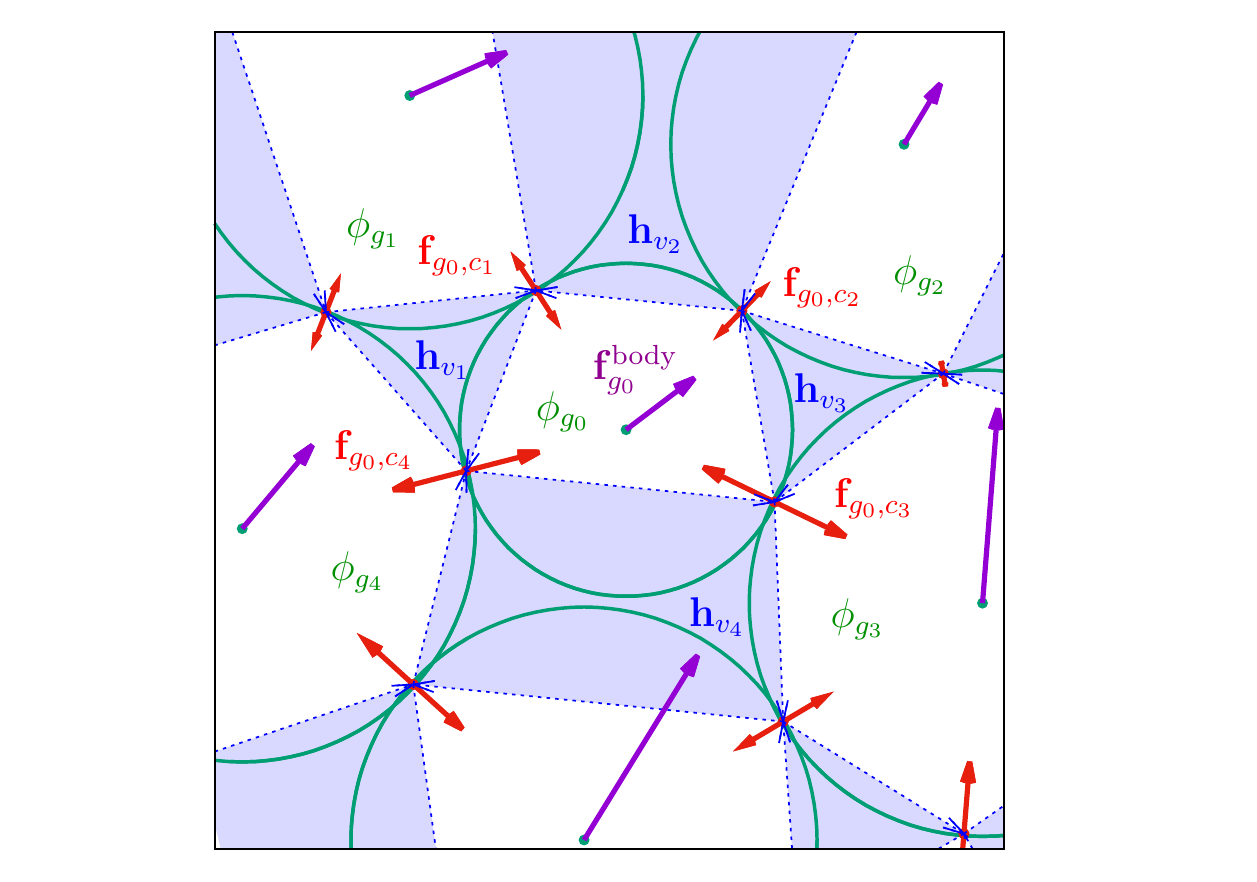}
\caption{The configuration of Fig. \ref{explanatory_figure1}, with external forces on the grains. To deal with the general case where body forces $\{\vec{f}^{\rm body}_g\}$ act on each grain (represented with purple arrows), we introduce auxiliary fields $\{\vec{\phi}\}$ that are associated with the grains ${g}$. The contact forces are now given by the difference in height variables and a difference of auxiliary fields. For example $\vec{f}_{g_0,c_1} = \vec{h}_{v_1} - \vec{h}_{v_2} + \vec{\phi}_{g_1} -\vec{\phi}_{g_0}$.}
\label{explanatory_figure2}
\end{center}
\end{figure} 


\subsection{Auxiliary Fields}
In order to account for the general case where contact forces and body forces are present within the system, we introduce an additional {\it auxiliary vector field} $\{ \vec{\phi}_g\}$ associated with the grain polygons. This serves to generalize the height construction of Eq. (\ref{height_definition}). Since each contact has two associated grains and two associated voids, the contact forces are now given by the difference of height variables on the voids (as before) {\it in addition to} a difference of auxiliary fields on the grains as
\begin{equation}
\vec{f}_{g,c} = \vec{h}_{v'} - \vec{h}_{v} + \vec{\phi}_{g'} - \vec{\phi}_{g}.
\label{phi_definition}
\end{equation}
Here the contact $c$ is shared between the grains $g$ and $g'$ and $v$ and $v'$ are once again ordered cyclically around the centre of the grain $g$. It is easy to see that this definition satisfies Newton's third law (Eq. \ref{third_law}). In the absence of body forces, the $\vec{\phi}$ field is identically zero and the above definition reduces to Eq. (\ref{height_definition}).
As an illustrative example we consider the previous configuration (Fig. \ref{explanatory_figure1}) with grain $g_0$ which is in contact with four neighbours $g_{1,2,3,4}$ through the contacts $c_{1,2,3,4}$, with additional body forces $\{\vec{f}^{\rm body}_g\}$ acting on each grain (see Fig. \ref{explanatory_figure2}). The forces acting on the grain $g_0$ are given by 
\begin{align}
\begin{tabular}{c c c c}
    $\vec{f}_{g_0,c_1}$ & $=$ & $\vec{h}_{v_1} - \vec{h}_{v_2}$ & $ + \vec{\phi}_{g_1} -\vec{\phi}_{g_0}$,\\
    $\vec{f}_{g_0,c_2}$ & $=$ & $\vec{h}_{v_2} - \vec{h}_{v_3}$ & $ + \vec{\phi}_{g_2} -\vec{\phi}_{g_0}$,\\
    $\vec{f}_{g_0,c_3}$ & $=$ & $\vec{h}_{v_3} - \vec{h}_{v_4}$ & $ + \vec{\phi}_{g_3} -\vec{\phi}_{g_0}$,\\
    $\underbrace{\vec{f}_{g_0,c_4}}_{\sum_c \vec{f}_{g_0,c}  = -\vec{f}^{\rm body}_{g_0}}$ & $=$ & $\underbrace{\vec{h}_{v_4} - \vec{h}_{v_1}}_{\sum_{(v,v')} \vec{h}_{v} -  \vec{h}_{v'}  = 0}$ & $\underbrace{+ \vec{\phi}_{g_4} -\vec{\phi}_{g_0}.}_{\sum_{g'} \vec{\phi}_{g'} - \vec{\phi}_{g_0} = \square^2 \vec{\phi}_0}$\\  
  \end{tabular}
\label{modified_height_definition}
\end{align}

The sum of the forces on the grain $g_0$ sum to the negative of the body force $\vec{f}^{\rm body}_{g_0}$ on the grain due to the mechanical equilibrium condition (Eq. (\ref{force_sum_body})). The difference of height fields around the grain $g_0$ sum to zero as before. The summation on the right involving the $\vec{\phi}$ fields on the grains $g'$ is simply the {\it network Laplacian} defined, on grain $g_0$ as
\begin{equation}
\square^2 \vec{\phi_0}= \vec{\phi}_{g_1}+ \vec{\phi}_{g_2} + \vec{\phi}_{g_3} + \vec{\phi}_{g_4} - 4 \vec{\phi}_{g_0}.
\end{equation}
It is straightforward to show that this Laplacian equation is valid for every grain. It is important to note that both the body forces and the auxiliary fields in the above equation are associated with the grains.
In general, one can write the relationship between $\{\vec{f}^{\rm body}\}$ and $\{\vec{\phi}\}$ in a vectorial notation as
\begin{equation}
\square^2 | \vec{\phi} \rangle = - | \vec{f}^{\rm body} \rangle,
\label{basic_equation}
\end{equation}
where  $| \vec{\phi}\rangle$ represents the vector $(\vec{\phi}_{g_1}, \vec{\phi}_{g_2} , ... \vec{\phi}_{g_{N_G}})$ and $| \vec{f}^{\rm body} \rangle$ is the vector of body forces  $(\vec{f}^{\rm body}_{g_1}, \vec{f}^{\rm body}_{g_2} , ... \vec{f}^{\rm body}_{g_{N_G}})$. Here $N_G$ is the total number of grains in the system. The Laplacian operator $\square^2$ is now a matrix acting on these states. Eq. (\ref{basic_equation}) represents our main result, and we can use it to analyse the response of a mechanically stable configuration to imposed body forces.
Given a set of body forces and the contact network, we can {\it invert this equation} to obtain the auxiliary fields $\{ \vec{\phi}\}$. The changes in the contact forces that develop as a response to an applied body force is then simply given by the difference of the $\vec{\phi}$ field at each contact (Eq. (\ref{modified_height_definition})).

\begin{figure} 
\begin{center}
\includegraphics[width=0.6\linewidth]
{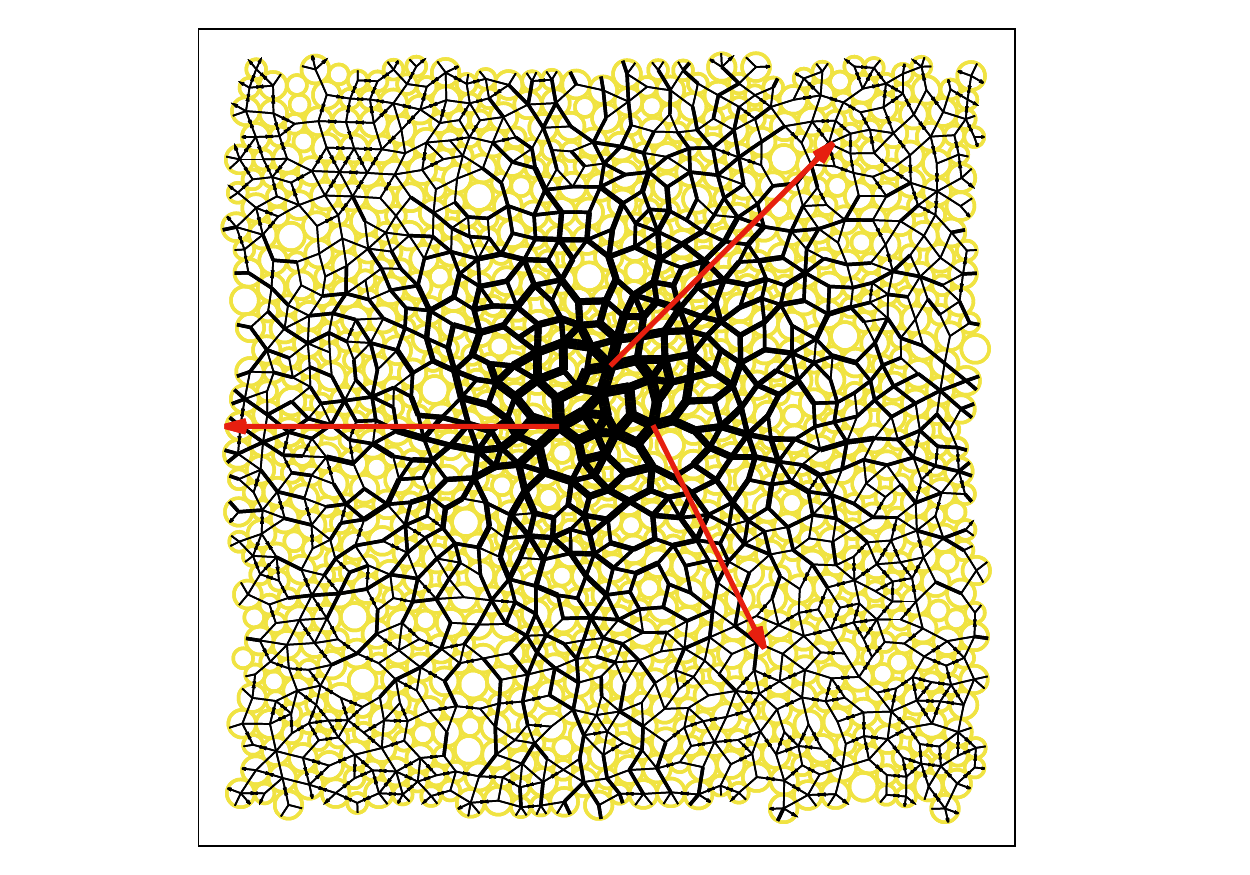}
\caption{The response of a system of soft disks to applied body forces (represented by red arrows). The system is prepared with periodic boundary conditions and uniform global compression. The magnitude of the change in contact forces are respresented by the thickness of the lines (in log scale) connecting the disks. The imposed body forces at the center of the system are $10^{-6}$ times smaller than the average contact force in the system. The sum of body forces on the system is zero. The inhomogeneous nature of the stress response is clearly illustrated.}
\label{response_full_figure}
\end{center}
\end{figure} 


Once we have determined the $\vec{\phi}$ field, we have local fields that incorporate the effect of the body forces at each contact. We can subtract the difference of these $\vec{\phi}$ fields, from the original contact forces to obtain {\it ``effective'' contact forces} ${\vec{\tilde{f}}}_{g,c}$ which satisfy the constraints of local force balance (Eq. (\ref{force_sum_zero})) and Newton's third law (Eq. (\ref{third_law})) with
\begin{equation}
\vec{\tilde{f}}_{g,c} = \vec{f}_{g,c} - \square \vec{\phi}.
\label{effective_forces}
\end{equation}
Here $\square \vec{\phi} = \vec{\phi}_{g'} - \vec{\phi}_{g}$ represents the gradient of the $\vec{\phi}$ field on the network.
These effective forces can then be used to determine the height fields on the voids, using the construction described in the Section \ref{height_section}.

In the case where a body force is imposed on a granular packing already in mechanical equilibrium, the difference between the effective contact forces and the original ones represents the response of the granular packing to this perturbation. In the most general case, these new effective forces would induce changes in the real space network to satisfy the other constraints of mechanical equilibrium:  torque balance and, in the case of frictional grains, the Coulomb condition of static friction. In many existing treatments of stress transmission and response in granular materials \cite{majumdar_coppersmith_pre_1996,bouchaud_review_2002}, structural changes are not allowed.  The argument being that for rough, rigid grains, there is an indeterminacy at the contact level that allows for multiple force configurations to be consistent with a given real-space contact network. Experiments have also analyzed responses to external forces that occur without any changes in the network \cite{geng_prl_2001}.  It should be remarked that allowing for network reorganization in response to external perturbations has been shown to have a significant  effect on the coarse-grained description of stress transmission\cite{goldenberg_nature_2005}.

\begin{figure} 
\begin{center}
\includegraphics[width=0.6\linewidth]
{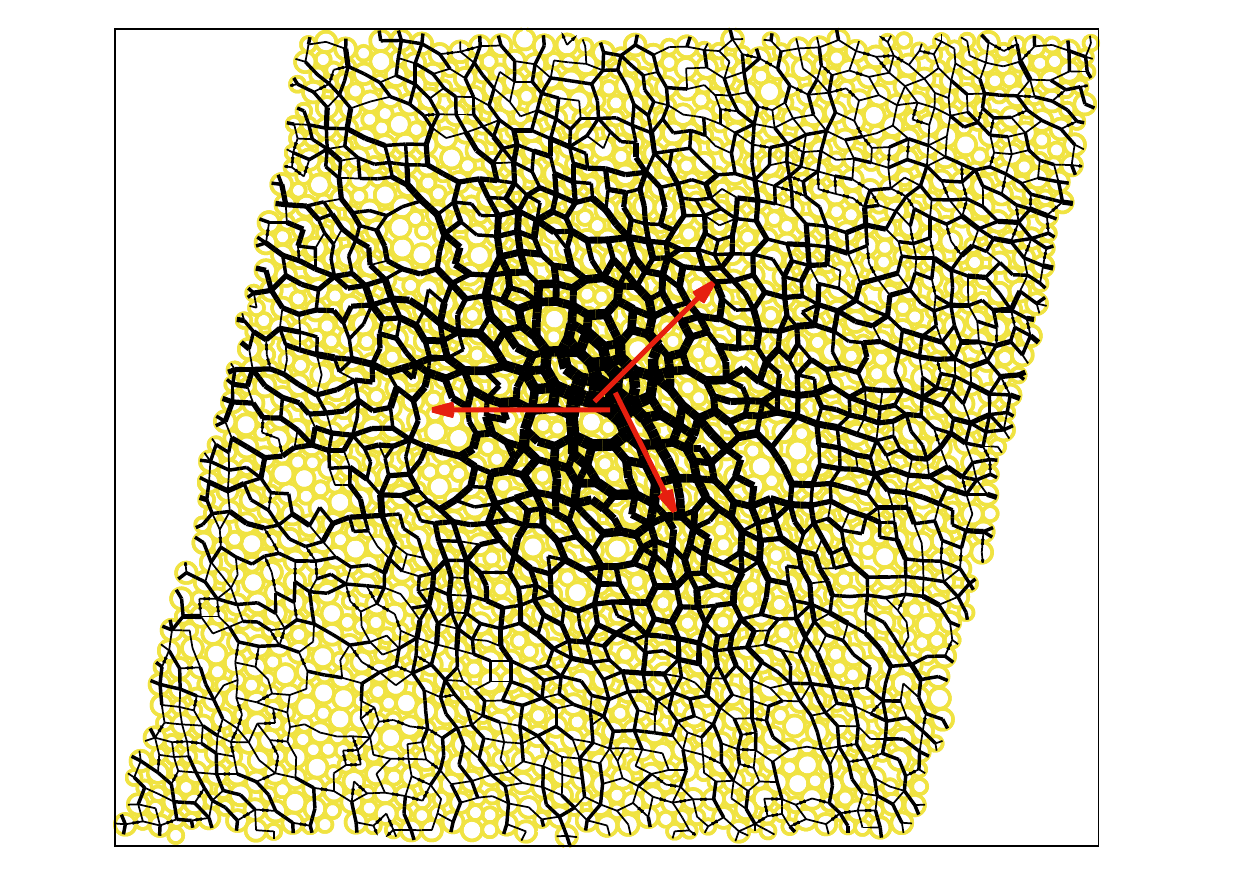}
\caption{The response of a sheared system of soft frictional disks to applied body forces (represented by red arrows). The magnitude of the change in contact forces are respresented by the thickness of the lines (in log scale) connecting the disks. The imposed body forces at the center of the system are $10^{-6}$ times smaller than the average contact force in the system. The sum of body forces on the system is zero. The system is prepared with Lees-Edwards boundary conditions with a global shear of $\gamma = 0.43$ \cite{vinutha_unpublished}. The response provides characteristic signatures of the emergence of ``force chains" along the compressive direction.}
\label{shear_response_full_figure}
\end{center}
\end{figure} 


In this paper, we focus on the solutions for the effective contact forces and their localization properties in the absence of rearrangements of the contact network.
For frictionless grains, which obey a given force law, a change in the forces in a given packing is necessarily accompanied by a {\it structural change}. For disks interacting via one-sided linear spring potentials, these real-space displacements can be obtained in closed form. In this case, the change in $\vec{\phi}$ fields directly represents the change in displacements of the particles, satisfying the required additional constraints of Section \ref{additional_constraints_section}.


\subsection{Inverting the Body Forces}

The network Laplacian is the adjacency matrix of the graph representing the contact network with an added diagonal matrix whose entries are the number of contacts of the grain corresponding to that row, and has several well known properties. In our case the network is a disordered planar graph. $\square^2$ has the eigenfunction expansion
\begin{equation}
\square^2 = \sum_{i=1}^{N_G}  \lambda_i | \lambda_i \rangle \langle \lambda_i |,
\label{eigenfunction_expansion}
\end{equation}
with $N_G$ being the dimensionality, i.e. the number of grains in the system. $\square^2$ has {\it one} zero eigenvalue, with eigenvector
\begin{equation}
\lambda_1 = 0, ~~| \lambda_1 \rangle = (1 1 1 ... 1).
\end{equation}
This trivial zero mode is a consequence of  the ``conservation'' law: $\sum_j \square^2_{i,j} =0$ for every row $i$ of the matrix representing the network Laplacian.
The rest of the eigenvalues are {\it all negative}.
We can next use the above eigenvalue expansion to invert the body forces and obtain the $\vec{\phi}$ fields in Eq. (\ref{basic_equation}). We define a restricted inverse $(\square^2)^{-1}$ of the Laplacian operator by projecting out the zero mode. We then have
\begin{equation}
\underbrace{\left( \sum_{i>1}  \frac{1}{\lambda_i} | \lambda_i \rangle \langle \lambda_i | \right)}_{(\square^2)^{-1}} \square^2 = \mathbb{I} - | \lambda_1 \rangle \langle \lambda_1 |.
\label{projected_inverse}
\end{equation}
Next, using Eq. (\ref{projected_inverse}) in Eq. (\ref{basic_equation}), we obtain the inversion
\begin{eqnarray}
\nonumber
- (\square^2)^{-1} | \vec{f}^{body} \rangle = | \vec{\phi} \rangle -|  \lambda_1 \rangle \langle \lambda_1 | \vec{\phi} \rangle \\
=| \vec{\phi} - \frac{1}{N}\sum_{i=1}^{N} \vec{\phi} ~~\rangle.
\end{eqnarray}

\subsection{Centre of Mass Frame}
The inversion of the above equations is more natural when one considers the center of mass frame of reference. We can define
\begin{eqnarray}
\nonumber
| \vec{\mathcal{F}}^{body} \rangle = | \vec{f}^{body} \rangle -|  \lambda_1 \rangle \langle \lambda_1 | \vec{f}^{body} \rangle \\
=| \vec{f}^{body} - \frac{1}{N}\sum_{i=1}^{N} \vec{f}^{body} \rangle.
\end{eqnarray}
Here $| \vec{\mathcal{F}}^{body} \rangle $ represents the vector of body forces in the center of mass frame.
This then leads to a more symmetric formulation of Eq. (\ref{basic_equation})
\begin{equation}
|\vec{\mathcal{F}}^{body} \rangle = -\square^2 | \vec{\phi} \rangle,
\end{equation}
along with the inversion equation
\begin{equation}
| \vec{\phi} \rangle =  - (\square^2)^{-1}  | \vec{\mathcal{F}}^{body} \rangle.
\end{equation}
The above equations provide a {\it unique} solution to $| \vec{\phi} \rangle$ for a given network and a given set of body forces.  If this solution fails to satisfy the other constraints of mechanical equilibrium such as torque balance, the network has to necessarily rearrange and indicates an instability of the network to this perturbation.  Our current treatment, which focuses only on force balance, cannot address these questions of instability. In the figures illustrating the inhomogeneous response (Figs. \ref{response_full_figure} and \ref{shear_response_full_figure}), we have used body forces that are much smaller than the average force (and therefore the overlaps) between the grains, leading to very small changes in the contact forces, keeping the connectivity of the network unperturbed in the process.

As an illustrative example of the stress response within this framework, in Fig. \ref{response_full_figure} we plot the changes in contact forces that develop as a response to localized body forces in a jammed packing of soft frictionless disks.
The body forces (represented by red arrows) act at the centers (i.e. centers of mass) of three grains separated by a small distance. This illustrates the effect of a localized stress perturbation to the packing. The reason for perturbing three grains is to create a non-trivial local perturbation that leaves the entire system in force balance. The changes in the contact forces in response to these body forces are obtained by solving Eq. (\ref{basic_equation}) for a given initial jammed packing. The inhomogeneous nature of the stress response is clearly illustrated. As a more dramatic example, we plot this response for a sheared packing of soft frictional disks \cite{vinutha_unpublished} in Fig. (\ref{shear_response_full_figure}). The response provides characteristic signatures of the emergence of ``force chains" in this case.    To construct these responses,  numerically simulated packings of frictionless and frictional grains were used to construct the network Laplacian.  The above set of equations were then used to calculate $| \vec{\phi} \rangle$ for a network, which was then used to calculate the changes in the contact forces resulting from  the imposed $| \vec{\mathcal{F}}^{body} \rangle.$  Below, we consider the relation of the response of an ensemble of packings to the spectral properties of an ensemble of graph Laplacians. 

\section{Spectrum of the Laplacian}

We note that our formulation is closely related to the diffusion equation for stresses obtained by going to the continuum limit of the $q$-model describing stress transmission in a granular pile created under gravity~\cite{bouchaud_review_2002}. 
In that formalism, the disorder is represented by the distribution of the $q_{i,j}$s that specify how the weight of a grain is split between different contacts.  This disorder renormalizes the diffusion constant in the equation for the Greens function for stress propagation. Stresses transmit essentially vertically downwards with a small spread \cite{bouchaud_review_2002}. The network Laplacian formulation is able to take care of the underlying randomness in the system through the disorder in the contact network. This framework also demonstrates that the $q_{i,j}$s cannot be considered as independent, random stochastic variables.  Instead, these variables that specify the redistribution of body forces are determined by the underlying network with its associated randomness.

\subsection{Stress Localization}

\begin{figure} 
\begin{center}
\includegraphics[width=0.45\linewidth]
{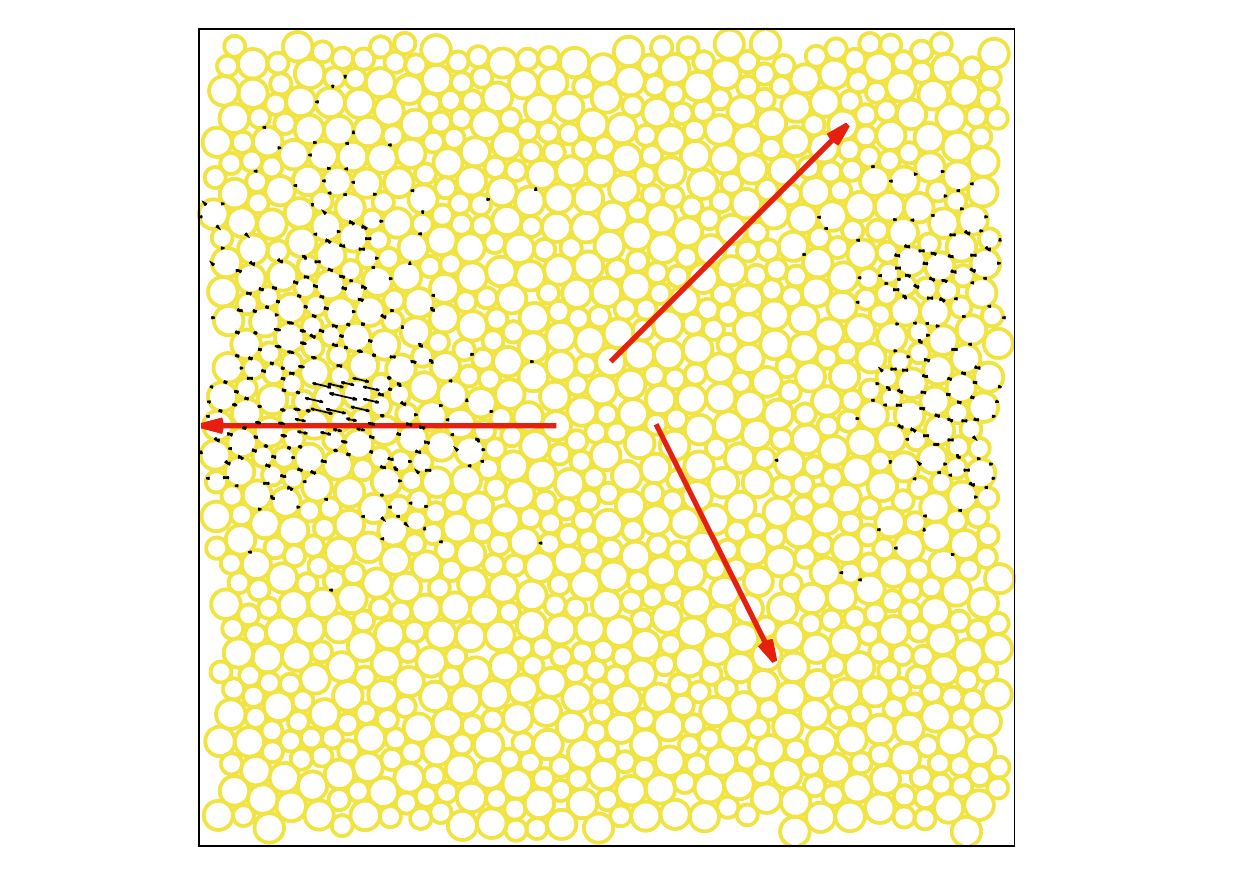}
\includegraphics[width=0.45\linewidth]
{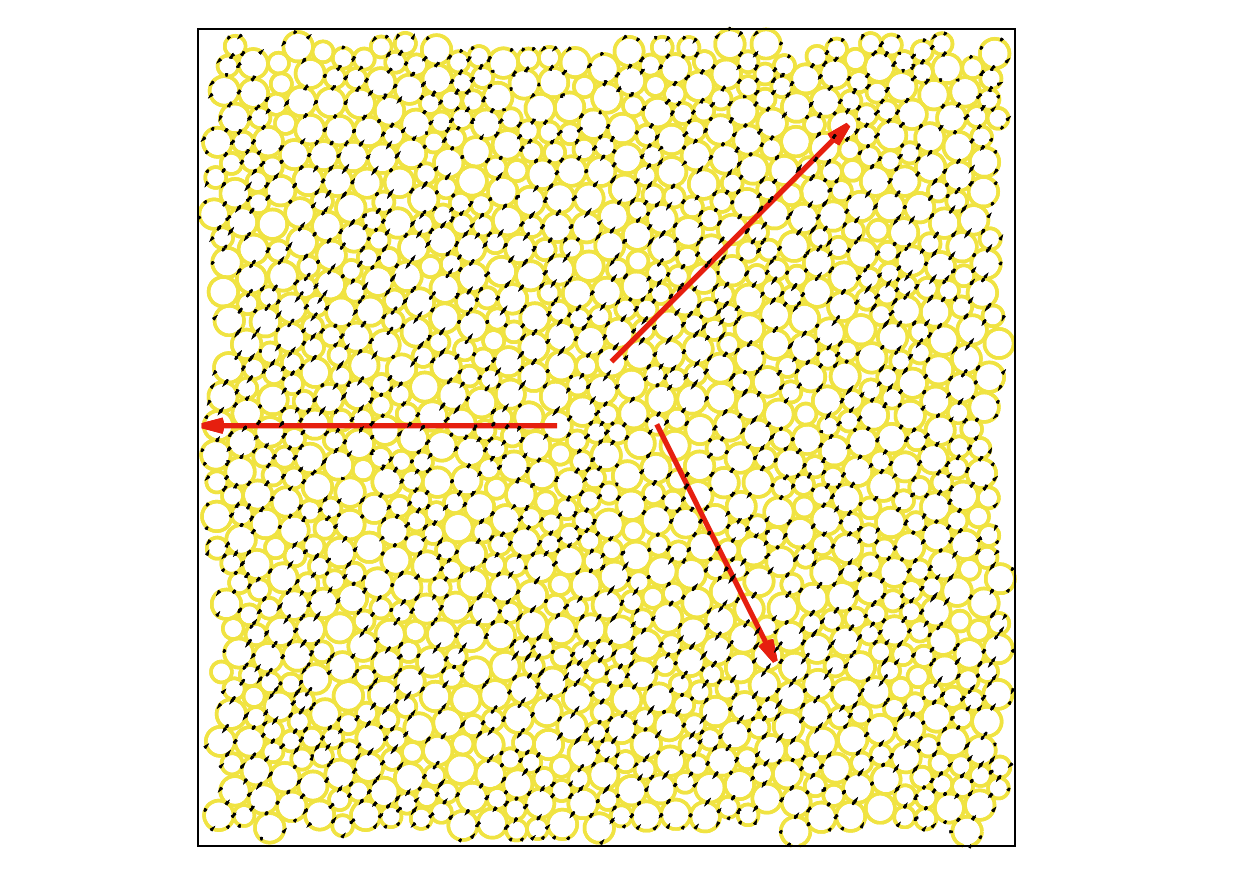}
\caption{The response of the system shown in Fig \ref{response_full_figure}, ({\bf left}) using only the largest negative eigenvector of the Laplacian matrix, illustrating a localized response, and ({\bf right}) using only the smallest negative eigenvector of the Laplacian matrix, illustrating a delocalized response. The black arrows represent the changes in the contact force vectors in response to the imposed body forces (red arrows).}
\label{response_eigenvalues_figure}
\end{center}
\end{figure} 

The connection to the problem of diffusion in the gauge potential formulation is immediately obvious as diffusion on the random planar graph is described precisely by Eq. (\ref{basic_equation}). In addition,
the discrete Laplacian matrix is important in many physical situations and several physical models on graphs rely heavily on it. For example, the kinetic energy term for hopping models, the 
dynamics of random vibrational networks, ferromagnetic $O(n)$ models as well as models of non-interacting bosons on graphs invoke the network Laplacian \cite{hastings_prl_2003,burioni_prl_2000}.
Several intriguing connections to the problem of localization follow immediately. We can therefore use the already sophisticated machinery developed in the field of localization to study the phenomenon of force localization in granular systems.

A natural question to consider is then the contribution of the different  eigenvectors of the Laplacian (Eq. (\ref{eigenfunction_expansion})) to the stress response of the system. Fig. \ref{response_full_figure} provided an illustration of the {\it total} change in contact forces as a response to imposed body forces for a single configuration of jammed frictionless disks. In Fig. \ref{response_eigenvalues_figure}, we plot the response of the contact forces for the same configuration using just the largest negative and the smallest negative eigenvalues of the network Laplacian. We find clear signatures of localization in the higher end of the spectrum. To better understand these localization properties, we study packings of soft disks numerically.

\subsection{Matrix Ensemble}
It is clear from the formulation in Section \ref{response_to_perturbation_section}, that the spatial localization, or lack thereof, of the $\vec{\phi}$ fields is determined by the nature of the eigenfunctions of the network Laplacian. The eigenvectors that possess a large overlap with $| \vec{\mathcal{F}}^{body} \rangle $ will contribute overwhelmingly to this response, and therefore it is important to understand how localized the eigenvectors of the Laplacian matrix are. 
In usual two dimensional localization problems, the disorder is manifest in the system as a quenched randomness, either explicitly in the interactions or in the spatial motion, which is then averaged over. In the case of granular packings, the disorder in the network of a given configuration plays the role of such a quenched variable. The ensemble of adjacency matrices can be thought of as an ensemble of random matrices with the randomness entering through the connectivities of the particles. The extent of stress localization is thus controlled by the ensemble of random matrices that represent the network Laplacians of disordered granular packings created through some protocol, and the nature of the perturbation.

%

\section{Numerical Simulations of Frictionless Grains}

In this section, we probe the spatial localization properties of the stress response in frictionless granular media using numerically simulated packings. We emphasize that our framework is not restricted to frictionless packings, we use frictionless packings as a model system to illustrate the application of our theoretical framework.
To do this, we consider the paradigmatic example of a system of frictionless soft disks interacting via one sided linear spring potentials. To study the stress response of this system, we create a jammed packing of soft frictionless disks and perturb this with spatially localized body forces \cite{ellenbroek_pg_2005}. We then measure the statistics of the forces that develop as a consequence of the imposed body forces using the formulation developed in Section \ref{response_to_perturbation_section}.

\begin{figure}
\begin{center}
\includegraphics[width=0.6\linewidth]
{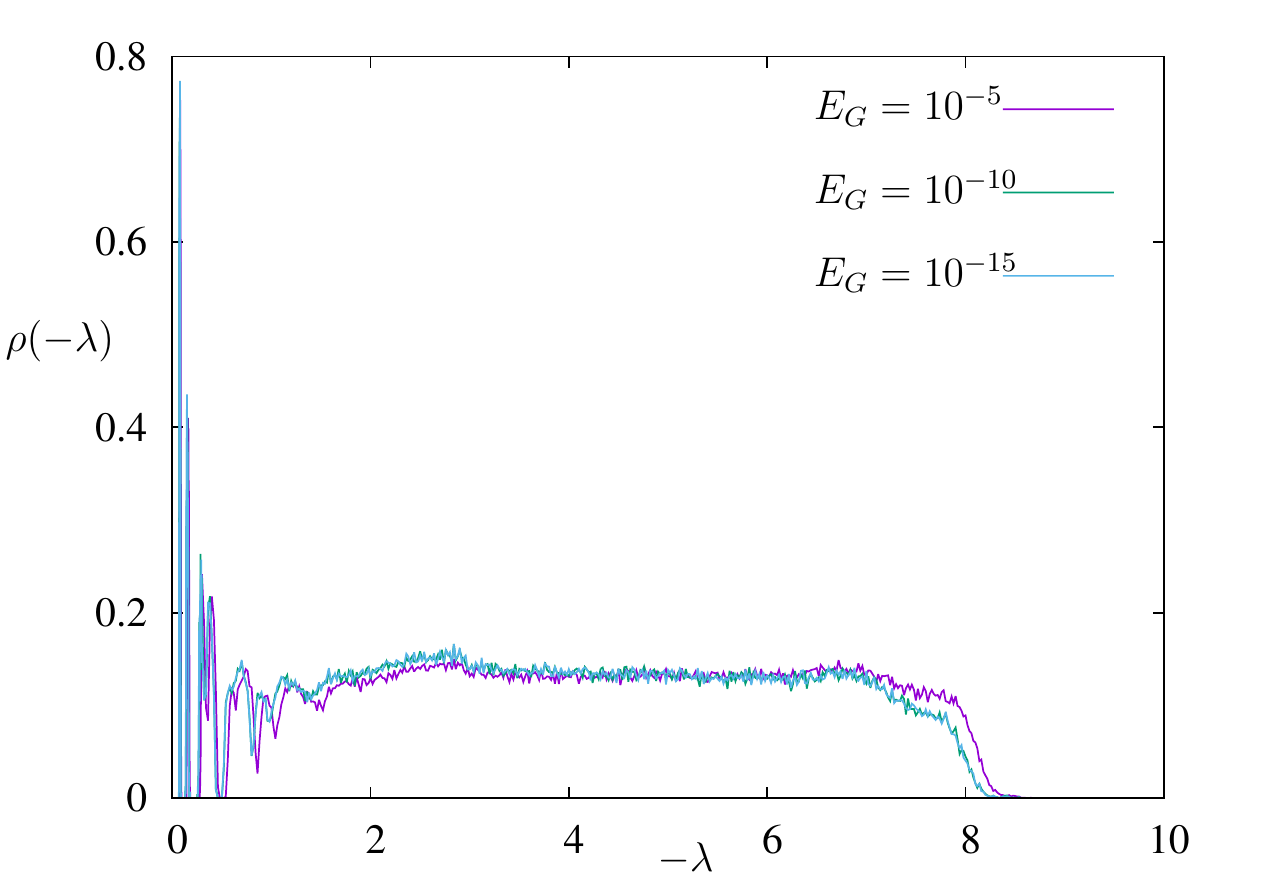}
\caption{The density of states $\rho(\lambda)$ of the eigenvalues $\lambda$ of the Laplacian matrix, for $N_G = 1024$ grains at different global energies ($E_G$). The data is averaged over $5000$ configurations.}
\label{dos_figure}
\end{center}
\end{figure}

\subsection{Jammed Packings of Soft Disks}
We simulate a system of soft disks interacting via linear spring potentials of the form
\begin{equation}
V(\vec{r}_{g, g'}) = \frac{1}{2} \left(1-\frac{|\vec{r}_{g,g'}|}{\sigma_{g,g'}}\right)^2 \Theta\left(1-\frac{|\vec{r}_{g,g'}|}{\sigma_{g,g'}}\right),
\label{potential_energy_functional}
\end{equation}
where $\Theta$ is the Heaviside function and $\sigma_{g, g'} = \sigma_{g} + \sigma_{g'}$ is the sum of the undistorted radii of disks $g$ and $g'$. 
The total energy per grain of the system is given by
\begin{equation}
E_G = \frac{1}{N_G} \sum_{(g,g')} V(\vec{r}_{g, g'}),
\label{energy_of_system}
\end{equation}
where the sum is taken over all pairs $(g,g')$, with $g \ne g'$. $N_G$ is the total number of grains in the packing.
We create jammed packings in mechanical equilibrium using a conjugate gradient minimization of Eq. (\ref{energy_of_system}). The number of grains that are part of the rigid structure of the contact network varies between different configurations, i.e. $N_G \equiv N_G - N_R$, where $N_R$ is the number of ``rattlers", particles that are not in contact with any of the others. This crucially decreases the dimensionality of the Laplacian matrix, making it singular, and therefore rattlers need to be removed from the system before any numerical procedure is implemented. We simulate systems of particles with a $50:50$ mixture of disks with diameter ratios $1:1.4$, at varying global energies between $E_G = 10^{-15}$ and $10^{-5}$ \cite{ramola_chakraborty_jstat_2016}. The number of grains in our simulations vary between $N_G = 512$ and $2048$.

\begin{figure}
\begin{center}
\includegraphics[width=0.6\linewidth]
{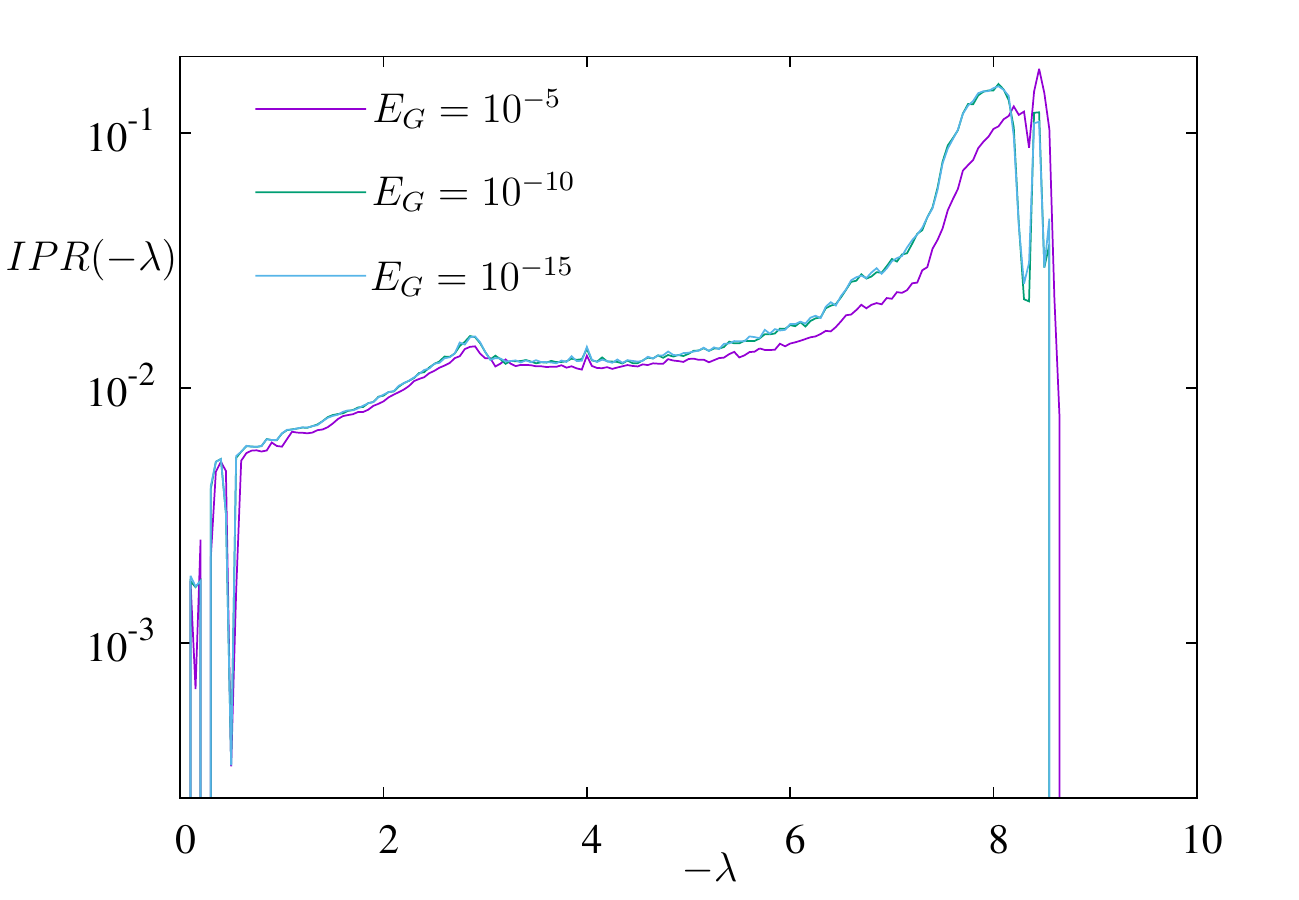}
\caption{The inverse participation ratio (IPR) of the Laplacian eigenvectors, for $N_G = 1024$ grains at different global energies ($E_G$). The low modes are delocalized whereas a large part of the spectrum is localized. The data is averaged over $5000$ configurations.}
\label{ipr_figure}
\end{center}
\end{figure}

\subsection{Exact Diagonalization}

We next exactly diagonalize the Laplacian matrix associated with the contact network of each configuration and measure the statistics of their eigenvalues and eigenvectors. We measure two characteristic signatures of localization in our system, namely, the density of states of the eigenvalues, and the inverse participation ratio \cite{slanina_epj_2012,clark_maestro_arxiv_2015}. The network Laplacian is a $N_G \times N_G$ real symmetric matrix with eigenvalues $\lambda_i, i = 1, . . . ,N_G$ and corresponding normalized eigenvectors $|\lambda \rangle \equiv ( e_{1, \lambda}, e_{2, \lambda}, ...,  e_{N_G, \lambda} )$. 
We measure the density of states $\rho(\lambda)$ of the eigenvalues of the Laplacian at different global energies. The density of states for an ensemble of $N_G = 1024$ grains at different global energies is illustrated in Fig. \ref{dos_figure}. We find single isolated states within the lower spectrum of eigenvalues and a continuum of states with higher eigenvalues.

We next compute the inverse participation ratios of the eigenvalues for different energies. The Inverse Participation Ratio (IPR) corresponding to an eigenvector is defined as
\begin{equation}
q^{-1}(\lambda) = \sum_{j} e_{j, \lambda}^4
\label{ipr_definition}
\end{equation}
For a localized mode the IPR would be of $O(1)$ and for a delocalized mode this quantity would be of $O(1/N_G)$. The IPR for an ensemble of $N_G = 1024$ grains at different global energies is illustrated in Fig. \ref{ipr_figure}. We find that a large part of the spectrum is in fact localized, with a small number of delocalized modes. This is similar to what one would expect in two dimensional disordered models \cite{abrahams_review_2010}, where states are in general localized. However, the nature of the disorder in granular systems still remains to be elucidated. This would require a detailed study of the properties of the random networks that arise in frictional granular packings, and is a fruitful direction for future research.


\section{Force Tiles with Body Forces}
\label{force_tile_section}

\begin{figure}
\begin{center}
\includegraphics[width=0.27\linewidth]{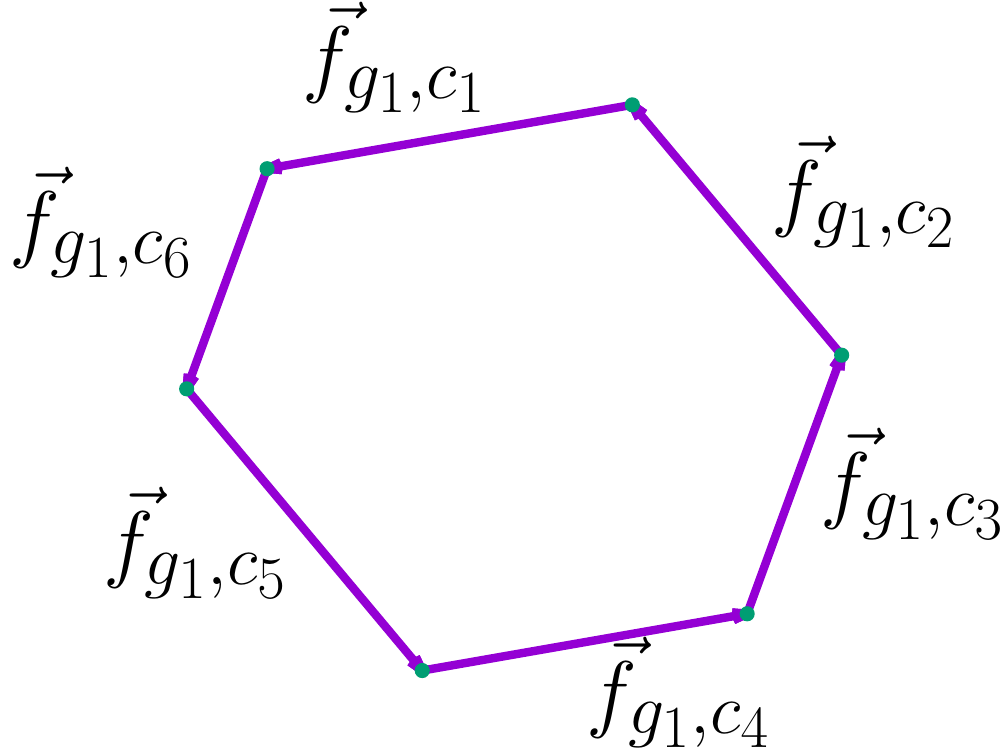}
\hspace{1cm}
\includegraphics[width=0.4\linewidth]{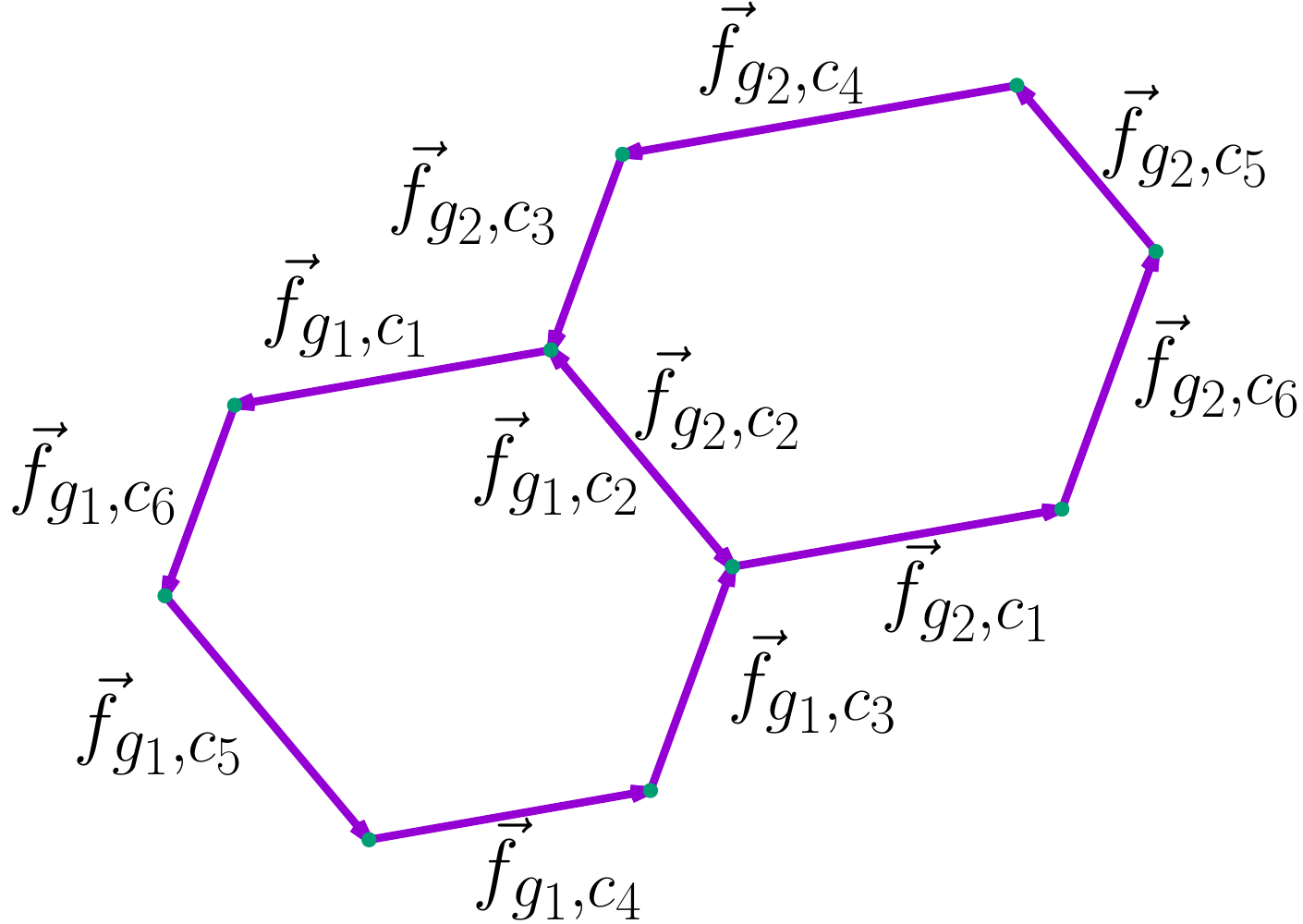}
\end{center}
\caption{({\bf Left}) Force polygon associated with a single grain $g_1$. The forces associated with the contacts of the grain, ordered cyclically, are arranged in a ``vector sum" (head to tail) forming a closed polygon. For frictionless systems, the normality of the forces ensures the convexity of these polygons. ({\bf Right}) Two force polygons line up along the equal and opposite contact forces $\vec{f}_{g_1,c_2} = -\vec{f}_{g_2,c_2}$. Iterating this procedure leads to the construction of a force tiling for the forces of the entire system.}
\label{force_polygons}
\end{figure}

In this section we discuss another application of the gauge potential framework developed in this paper, namely the construction of ``force tiles" for systems with body forces.
The condition of mechanical equilibrium associated with a static granular system with {\it purely pairwise contact forces} allows one to construct a useful representation of the forces in the system known as a force tile. This is constructed as follows. Using Eq. (\ref{force_sum_zero}), the ``vector sum" of the forces on each grain, i.e. the force vectors associated with the contacts of each grain arranged head to tail (with a cyclic convention), form a {\it closed polygon}. This forms a ``force polygon" associated with this grain (see Fig. \ref{force_polygons}). Since the sum is taken cyclically over the contacts for each grain, we obtain {\it convex polygons} for frictionless systems. In frictional systems, the force polygons can be non convex and even self-intersecting making the graph non planar. Next, Eq. (\ref{third_law}) imposes the condition that every force vector in the system, has an equal and opposite counterpart that belongs to its neighboring grain. This leads to the force polygons being exactly edge-matching, and one can then use this fact to tile these polygons next to each other (see Fig. \ref{force_polygons}). This construction  produces a network known as the {\it force tile network}, with the edges representing the forces in the system. 
The two representations: force tilings and height fields defined in Section \ref{height_section}
are related. It is easy to see that the positions of the vertices of the force tiles represent the height vectors starting from an arbitrary origin, since the vector distance between these vertices provides the forces (in correspondence with the definition of the heights). Since each face of the force tiling graph is uniquely associated with a grain, the adjacency of the faces is the adjacency of the grains in a packing. The adjacency of the vertices are simply the adjacency of the voids (since the heights are associated with the void polygons).
The force tiling representation has several intriguing properties \cite{tighe_jstat_2011}, an important one being that the distances between the vertices represents a measure of the amount of stress between two points in a packing. The extent of the tiling provides the total amount of stress in the packing. Since force tilings provide information about the nature of the stress distribution within a system, they provide sensitive measures with which to characterize stress induced transitions in granular media \cite{sarkar_prl_2013,sarkar_pre_2016}. In the presence of body forces, however, the force tile construction fails  \cite{tighe_pg_2009}. This is because the force polygons no longer close as the contact forces do not sum precisely to zero. This makes constructing force tiles for granular piles and for systems with hydrodynamic drag impossible.

\begin{figure}
\begin{center}
\includegraphics[width=0.65\linewidth]{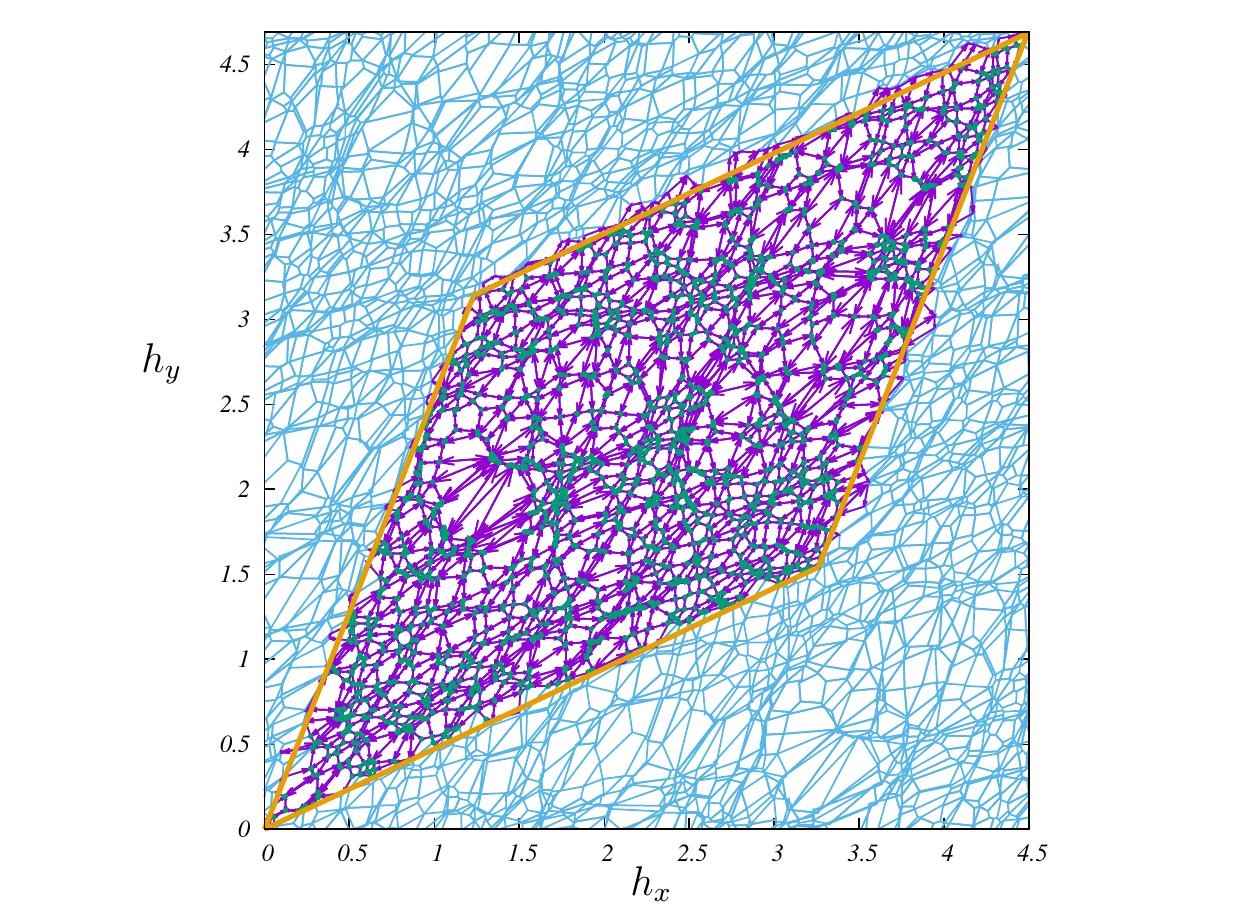}
\caption{The force tiling associated with a dense suspension of $2000$ soft disks with pairwise hydrodynamic interactions and high drag forces. The system is prepared with Lees-Edwards boundary conditions with a controlled shear rate \cite{mari_unpublished}. The viscous drag acts as a body force on each grain, the procedure in Eq. (\ref{basic_equation}) allows us to uniquely determine the positions of the vertices of the tiling (green dots), up to a global translation. These positions correspond to the values of the height field $\{\vec{h} \equiv (h_x, h_y)\}$ defined on the voids. The forces (purple arrows) are normalized by the average contact force in the system. The extent of the tiling (orange box) represents the total amount of stress in the system. The blue regions represent periodic copies of the system.} 
\label{bodyforce_tile_figure}
\end{center}
\end{figure}

The Laplacian framework developed in this paper can be used to extend the construction of force tiles for systems where contact forces are not the only forces in the system. Given the contact and body forces in the system, by using the network of contacts one can construct the $\vec{\phi}$ field as detailed in Section \ref{response_to_perturbation_section}. This then allows for a unique construction of the height field $\{\vec{h}\}$ which are the vertices of the force tiling. As an illustrative example, in Fig. \ref{bodyforce_tile_figure} we show the force tiling associated with a dense suspension of $2000$ soft disks with pairwise hydrodynamic interactions and high drag forces \cite{mari_unpublished}. In this case, the viscous drag acts as a body force on each grain. The positions of the vertices were computed by inverting Eq. (\ref{basic_equation}).

\section{Discussion}
In this paper we have discussed how stresses in granular packings are transmitted through the network of contact forces in response to external perturbations. Our formalism allowed us to construct force-balanced solutions on disordered networks that respect {\it vector} force balance at the microscopic level. This new potential formulation opens several interesting avenues. The Laplacian framework can be easily extended to systems where contact forces are not the only forces present in the system. This occurs frequently in systems with hydrodynamic forces, where viscous drag plays a major role, and granular piles in a gravitational field.

Our construction assumes that there is enough indeterminacy in the forces at the contact level  depending on how a packing is created. This is particularly true for systems with frictional grains where an exact force law is not applicable.  Even so, the constraints that are left out of the analysis such as the Coulomb constraint and torque balance would change the effect of the response. If we assume, as in the $q$-model,   that  roughness at the grain level would lead to an indeterminacy in the actual position of the contacts, then one can find a torque balanced solution as long as the perturbation is small enough, and the network does not rearrange. Our  construction therefore accurately describes systems near the infinitely rigid limit for which there is a large separation scales between forces and displacements \cite{vanhecke_prl_2004}. Generalizations of the framework presented in this work to  account for the other  constraints  of mechanical equilibrium, and allowing for network rearrangements would be an interesting avenue for future research.

\section{Acknowledgements}
This work has been supported by NSF-DMR 1409093 and the W. M. Keck Foundation.



\addcontentsline{toc}{section}{\protect\bibname}
\begin{thebibliography}{10}
\bibitem{behringer_review_2016} R. P. Behringer, ``Forces in Static Packings.", {\it Handbook of Granular Materials} (CRC Press, NY, 2016).
\bibitem{majumdar_coppersmith_science_1995} C. H. Liu, S. R. Nagel, D. A. Schecter, S. N. Coppersmith, S. N. Majumdar, and T. A. Witten, {\it Force fluctuations in bead packs}, Science {\bf 269}, 513 (1995).
\bibitem{majumdar_coppersmith_pre_1996} S. N. Coppersmith, C. H. Liu, S. N. Majumdar, O. Narayan, and T. A. Witten, {\it Model for force fluctuations in bead packs}, Phys. Rev. E {\bf 53}, 4673 (1996).
\bibitem{mueth_jaeger_pre_1998} D. M. Mueth, H. M. Jaeger, and S. R. Nagel, {\it Force distribution in a granular medium}, Phys. Rev. E {\bf 57}, 3164 (1998).
\bibitem{cates_wittmer_prl_1998} M. E. Cates, J. P. Wittmer, J.-P. Bouchaud, and P. Claudin, {\it Jamming, force chains, and fragile matter}, Phys. Rev. Lett. {\bf 81}, 1841 (1998).
\bibitem{bouchaud_claudin_crp_2002} J.-P. Bouchaud, P. Claudin, E. Cl\'ement, M. Otto and G. Reydellet, {\it The stress response function in granular materials}, C. R. Physique {\bf 3}, 141 (2002).
\bibitem{geng_prl_2001} J. Geng, D. Howell, E. Longhi, R. P. Behringer, G. Reydellet, L. Vanel, E. Cl\'ement, and S. Luding, {\it Footprints in sand: the response of a granular material to local perturbations}, Phys. Rev. Lett. {\bf 87}, 035506 (2001).
\bibitem{erpelding_epl_2010} M. Erpelding, A. Amon, and J. Crassous, {\it Mechanical response of granular media: New insights from diffusing-wave spectroscopy}, Eur. Phys. Lett. {\bf 91}, 18002 (2010).
\bibitem{goldenberg_prl_2002} C. Goldenberg and I. Goldhirsch, {\it Force chains, microelasticity, and macroelasticity}, Phys. Rev. Lett. {\bf 89}, 084302 (2002).
\bibitem{goldenberg_nature_2005}  C. Goldenberg and I. Goldhirsch, {\it Friction enhances elasticity in granular solids}, Nature {\bf 435}, 188 (2005).
\bibitem{goldenberg_prl_2006} C. Goldenberg, A. P. F. Atman, P. Claudin, G. Combe, and I. Goldhirsch, {\it Scale separation in granular packings: stress plateaus and fluctuations}, Phys. Rev. Lett. {\bf 96}, 168001 (2006).
\bibitem{goldenberg_pre_2008} C. Goldenberg and I. Goldhirsch, {\it Effects of friction and disorder on the quasistatic response of granular solids to a localized force}, Phys. Rev. E, {\bf 77}, 041303 (2008).
\bibitem{yan_arxiv} L. Yan, J.-P. Bouchaud, and M. Wyart, {\it Edge mode amplification in disordered elastic networks}, arXiv:1608.07222 (2016).
\bibitem{reydellet_Clement_PRL2001} G. Reydellet, and E. Cl\'ement, {\it Green's function probe of a static granular piling}, Phys. Rev. Lett. {\bf 86}, 3308 (2001).
\bibitem{atman_epje_2005} A. P. F. Atman, P. Brunet, J. Geng, G. Reydellet, P. Claudin, R. P. Behringer, and E. Cl\'ement, {\it From the stress response function (back) to the sand pile ``dip''}, Eur. Phys. J E {\bf 17}, 93 (2005).
\bibitem{atman_jphyscm_2005} A. P. F. Atman, P. Brunet, J. Geng, G. Reydellet,
G. Combe, P. Claudin, R. P. Behringer, and E. Cl\'ement, {\it Sensitivity of the stress response function to packing preparation},
J. Phys. Cond. Mat. {\bf 17}, S2391 (2005).
\bibitem{atman_epl_2013} A. P. F. Atman, P. Claudin, G. Combe and R. Mari, {\it Mechanical response of an inclined frictional granular layer approaching unjamming}, Europhys. Lett. {\bf 101}, 44006 (2013).
\bibitem{gland_epje_2006} N. Gland, P. Wang and H.A. Makse, {\it Numerical study of the stress response of two-dimensional dense granular packings}, Eur. Phys. J. E {\bf 20}, 179 (2006).
\bibitem{bouchaud_epje_2001}
J.P. Bouchaud, P. Claudin, D. Levine and M. Otto, {\it Force chain splitting in granular materials: A mechanism for large-scale pseudo-elastic behaviour}, Eur.
Phys. J. E {\bf 4}, 451 (2001).
\bibitem{socolar_epje_2002} J. E. S. Socolar, P. Claudin, and D. G. Schaeffer, {\it Directed force chain networks and stress response in static granular materials}, Eur. Phys. J. E {\bf 7}, 353 (2002), and Erratum: Eur. Phys. J. E {\bf 8}, 453 (2002).
\bibitem{bouchaud_review_2002}
See J.-P. Bouchaud  in
  {\it Slow Relaxations and Nonequilibrium Dynamics in Condensed Matter},
  J.-P. Bouchaud, J.~L. Barrat, M.~Feigelman, J.~Kurchan, and J.~Dalibard, eds.,
  vol.~77, pp.~185.
\newblock Les Ulis: EDP Sciences, 2003.
\bibitem{narayan_pre_2000} O. Narayan, {\it Vector lattice model for stresses in granular materials}, Phys. Rev. E {\bf 63}, 010301 (2000).
\bibitem{tighe_pg_2009} B.P. Tighe, {\it Granular lattice models with gravity}, Powders and Grains 2009, American Institute of Physics, 305-308 (2009).
\bibitem{bouchaud_cates_claudin_jpI_1995}
J.-P. Bouchaud, M. E. Cates and P. Claudin, {\it Stress distribution in granular media and nonlinear wave equation}, J. Phys. I France {\bf 5}, 639 (1995).
\bibitem{vanel_howell_pre_1999} L. Vanel, D. Howell, D. Clark, R.P. Behringer and E. Cl\'ement, {\it Memories in sand: Experimental tests of construction history on stress distributions under sandpiles}, Phys. Rev. E {\bf 60}, R5040 (1999).
\bibitem{satake_mom_1993} M. Satake, {\it New formulation of graph-theoretical approach in the mechanics of granular materials}, Mechanics of Materials {\bf 16}, 65 (1993).
\bibitem{ball_blumenfeld_prl_2002} R. C. Ball and R. Blumenfeld, {\it Stress field in granular systems: loop forces and potential formulation}, Phys. Rev. Lett. {\bf 88}, 115505 (2002).
\bibitem{kadanoff_review_rmp_1999} L. P. Kadanoff, {\it Built upon sand: Theoretical ideas inspired by granular flows}, Rev. Mod. Phys. {\bf 71}, 435 (1999).
\bibitem{ramola_chakraborty_jstat_2016} K. Ramola and B. Chakraborty, {\it Disordered contact networks in jammed packings of frictionless disks}, J. Stat. Mech. 114002 (2016).
\bibitem{ramola_chakraborty_arxiv_2016} K. Ramola and B. Chakraborty, {\it Scaling Theory for the Frictionless Unjamming Transition}, Phys. Rev. Lett. {\bf 118}, 138001 (2017).
\bibitem{henkes_chakraborty_pre_2009} S. Henkes and B. Chakraborty, {\it Statistical mechanics framework for static granular matter}, Phys. Rev. E {\bf 79}, 061301 (2009).
\bibitem{degiuli_thesis} E. DeGiuli, {\it Continuum Limits of Granular Systems}, Ph.D. Thesis, University of British Columbia, (2013).
\bibitem{degiuli_pre_2011} E. DeGiuli and J. McElwaine, {\it Laws of granular solids: Geometry and topology}, Phys. Rev. E {\bf 84}, 041310 (2011).
\bibitem{vinutha_unpublished} H. Vinutha, private communication, the configuration in Fig. \ref{shear_response_full_figure} was produced using the LAMMPS software.
\bibitem{hastings_prl_2003} M. Hastings, {\it Random vibrational networks and the renormalization group}, Phys. Rev. Lett. {\bf 90}, 148702 (2003).
\bibitem{burioni_prl_2000} R. Burioni, D. Cassi and C. Destri, {\it $n \to \infty$  limit of $O(n)$ ferromagnetic models on graphs}, Phys. Rev. Lett. {\bf 85}, 1496 (2000). 
\bibitem{ellenbroek_pg_2005} W. Ellenbroek, E. Somfai, W. van Saarloos and M. van Hecke, {\it Force response as a probe of the jamming transition}, Powders and Grains 2005, p. 377, Taylor and Francis (2005).
\bibitem{slanina_epj_2012} F. Slanina, {\it Localization of eigenvectors in random graphs}, Eur. Phys. J. B {\bf 85}, 361 (2012).
\bibitem{clark_maestro_arxiv_2015} T. B. P. Clark and A. Del Maestro, {\it Moments of the inverse participation ratio for the Laplacian on finite regular graphs}, arXiv:1506.02048 (2015).
\bibitem{abrahams_review_2010} E. Abrahams (Ed.), {\it 50 Years of Anderson Localization}
(World Scientific, Singapore, 2010).
\bibitem{mari_unpublished} R. Mari, private communication.
\bibitem{tighe_jstat_2011}  B. P. Tighe and T. J. H. Vlugt, {\it Stress fluctuations in granular force networks}, J. Stat. Mech. P04002 (2011).
\bibitem{sarkar_prl_2013} S. Sarkar, D. Bi, J. Zhang, R. P. Behringer, and B. Chakraborty, {\it Origin of rigidity in dry granular solids}, Phys. Rev. Lett. {\bf 111}, 068301 (2013).
\bibitem{sarkar_pre_2016} S. Sarkar, D. Bi, J. Zhang, J. Ren, R. P. Behringer, and B. Chakraborty, {\it Shear-induced rigidity of frictional particles: Analysis of emergent order in stress space}, Phys. Rev. E {\bf 93}, 042901 (2016).
\bibitem{vanhecke_prl_2004} J. H. Snoeijer, T. J. H. Vlugt, M. van Hecke, and W. van Saarloos, {\it Force network ensemble: a new approach to static granular matter}, Phys. Rev. Lett. {\bf 92}, 054302 (2004).
\end{thebibliography}
\end{document}